\def   \ni {\noindent}

\def   \ssk {\vskip  5truept}

\def   \bsk {\vskip 15truept}
 
\def   \newpage {\vfill\eject}
\def   \newline {\hfil\break}

\documentstyle[epsfig]{article}
\begin{document}

\hsize 5truein
\vsize 8truein
\font\abstract=cmr8
\font\keywords=cmr8
\font\caption=cmr8
\font\references=cmr8
\font\text=cmr10
\font\affiliation=cmssi10
\font\author=cmss10
\font\mc=cmss8
\font\title=cmssbx10 scaled\magstep2
\font\alcit=cmti7 scaled\magstephalf
\font\alcin=cmr6 
\font\ita=cmti8
\font\mma=cmr8
\def\ref{\par\noindent\hangindent 15pt}
\null


\title{\ni X-Ray Spectral Behavior of the Relativistic Jet Source Cygnus X-3
}                                               

\bsk \bsk
\author{\ni M. L. McCollough$^{1}$, C. R. Robinson$^{1}$, S. N. Zhang$^{1}$, B. A. Harmon$^{2}$,
W. S. Paciesas$^{3}$, S. Dieters$^{3}$, S. Phengchamnan$^{3}$
R. M. Hjellming$^{4}$, M. Rupen$^{4}$, A. J. Mioduszewski$^{5}$,
E.~B. Waltman$^{6}$, F. D. Ghigo$^{7}$, G. G. Pooley$^{8}$,
R. P. Fender$^{9}$, W. Cui$^{10}$,
S. Trushkin$^{11}$
}                                                       
\bsk
\affiliation{ 1) Universities Space Research Association,
Huntsville, AL 35806, U.S.A.
}                                                
\affiliation{ 2) ES84 NASA/Marshall Space Flight Center, Huntsville, AL
      35812, U.S.A.
}                                                
\affiliation{ 3) University of Alabama in Huntsville, Huntsville, AL
      35899, U.S.A.
}                                                
\affiliation{ 4) National Radio Astronomy Observatory/VLA, Socorro,
      NM 87901, U.S.A.
}                                                
\affiliation{ 5) JIVE/National Radio Astronomy Observatory/VLA, Socorro,
      NM 87901, U.S.A.
}                                                
\affiliation{ 6) Naval Research Laboratory, Washington, D.C. 20375, 
      U.S.A.
}                                                
\affiliation{ 7) National Radio Astronomy Observatory/GBI, Green Bank,
      WV 24944, U.S.A. 
}                                                
\affiliation{ 8) Mullard Radio Astronomy Observatory, Cambridge, U.K.
}                                                
\affiliation{ 9) University of Amsterdam, Kruislaan 403, 1098 SJ 
      Amsterdam, The Netherlands
}                                                
\affiliation{ 10) Massachusetts Institute of Technology, Cambridge, MA 
      02139, U.S.A.
}                                                
\affiliation{ 11) Special Astrophysical Observatory, Nizhnij Arkhyz, 
      357147, Russia
}                                                
\bsk
\baselineskip = 12pt
\abstract{ABSTRACT \ni Cyg X-3 is an unusual X-ray binary which shows remarkable correlative behavior
between the hard X-ray, 
soft X-ray, 
and the radio. 
We present an
analysis of these long term light curves in the context of spectral changes of the system.  
This analysis will also incorporate a set of pointed 
RXTE observations made during a period when Cyg X-3 made a transition from a 
quiescent radio state to a flaring state (including a major flare) and then 
returned to a quiescent radio state.}  
\bsk
\baselineskip = 12pt
\keywords{\ni KEYWORDS: Cygnus X-3; X-Ray Binary; Radio Source; Relativistic Jets.
}               

\bsk
\baselineskip = 12pt


\text{\ni 1. INTRODUCTION
\ssk
\ni     
     Cyg X-3 is a very unusual X-ray binary  
(see Bonnet-Bidaud \& Chardin 1988 for a review) which does not 
fit well into any of the established classes of X-ray binaries. 
Cyg X-3 is also a very active radio source which has shown the presence of relativistic jets 
(Mioduszewski et al. 1998).

     In recent studies (McCollough et al. 1997a, 1997b, 1998) the 
following discoveries 
were made:  
{\bf (a)} During times of moderate radio brightness ($\sim 100$ mJy), and low 
variability the hard X-ray (HXR) flux anticorrelates with the radio.  
{\bf (b)} During periods of major flaring activity in the radio the HXR flux switches 
from an anticorrelation to a correlation with the radio.  
{\bf (c)} The HXR flux has been shown to anticorrelate with the soft X-ray (SXR).  This occurs in
both the low and high SXR states. 

\newpage

\bsk
\ni 2. LIGHTCURVES AND HARDNESS RATIOS 
\ssk
\ni

To better understand the behavior of Cyg X-3 in the X-ray, lightcurves and hardness ratios have been 
created from the CGRO/BATSE and RXTE/ASM data:



{\bf (a) BATSE Hardness Ratio:}  In Fig. 1 is the (50--100~keV)/(20--50~keV) hardness ratio as a function 
of the
20--100~keV flux.  One can see from this plot the HXR spectrum becomes harder during
flaring activity in the radio.  

{\bf (b) ASM Lightcurves/Hardness Ratios:} From the three ASM energy bands two hardness ratios can be 
created 
[$H_{low}$ = (3.0--4.8~keV)/(1.3--3.0~keV), $H_{high}$ = (4.8--12~keV)/(3.0--4.8~keV)].  
In Fig. 2 are time histories of the ASM flux and the hardness ratios.   
$H_{low}$ shows no trend with activity in the radio.  But $H_{high}$ becomes softer during
radio flaring and becomes harder during periods of radio quiescence.

{\bf (c) BATSE/ASM Hardness Ratio:} Fig. 3 is a (20--50~keV)/(4.8--12~keV) hardness ratio as a function of 
the total ASM flux.  The spectrum becomes softer as the ASM flux increases.


The anticorrelation between the HXR and SXR in Cyg X-3 indicates a pivoting behavior similar to that seen in 
Cyg X-1 (Zhang et al. 1997).  However, BATSE hardness ratios indicate that the spectrum above 20 keV 
becomes softer during
times of radio quiescence, which implies something more complicated than simple spectral pivoting.

\bsk
\ni 3. POINTED RXTE OBSERVATIONS 
\ssk
\ni 

To probe the various X-ray/radio states of activity in Cyg X-3 a series of 
target of opportunity observations, with RXTE, were made starting during an 
extended quenched radio
state and following Cyg X-3 through a large flare and into radio quiescent state. 

Cyg X-3's X-ray spectrum is known to be complicated with several different 
components 
(Nakamura et al. 1993).
The most dominant component is an absorbed power-law with an 
exponential cut-off at high energy (a Comptonized spectrum).
In Fig. 4 we shown three count spectra overlayed which  
show a pivoting around 10 keV.  During radio quiescent activity the spectrum hardens but above 15 keV the 
spectrum rolls over and becomes steeper, explaining the softening seen in the HXRs.
It can be seen
that the X-ray spectrum substantially changes for the different 
radio states in a way consistent with both ASM and BATSE 
hardness ratios.















}



\bsk
\baselineskip = 12pt


{\references \ni REFERENCES
\ssk

\ref Bonnet-Bidaud, J.M. \& Chardin, G. 1988, Physics Reports, 170, 326. 
\ref McCollough, M.L., et al. 1997a, in ``Transparent 
   Universe'', ESA, SP-382, p. 263. 
\ref McCollough, M.L., et al. 1997b, 4th Compton 
   Symposium, (AIP Press) p. 813.
\ref McCollough, M.L., et al. 1998a, ApJ, submitted. 
\ref Mioduszewski, A.J. et al. 1998, Proc. of IAU Colloquium 164, 351. 
\ref Nakamura, H., et al. 1993, MNRAS, 261, 353. 
\ref Zhang, S. N., et al. 1997, ApJ, 477, L95.
}

\newpage

\begin{figure}                                                                  
\centerline{\psfig{file=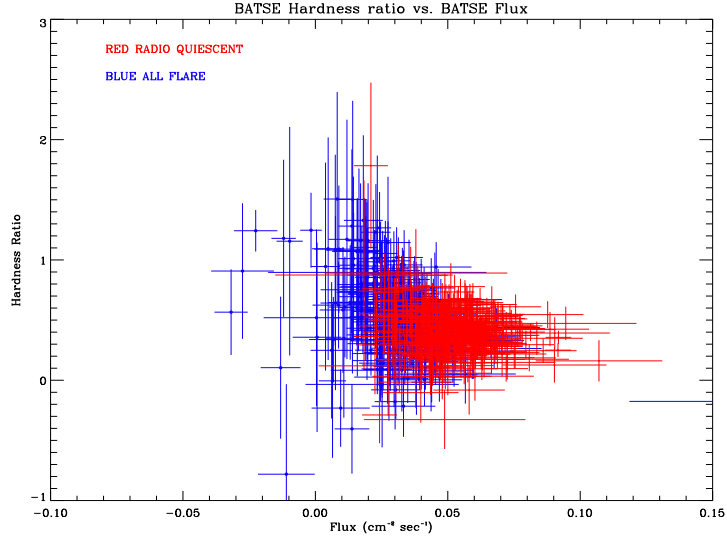,angle=90.0,height=7.25cm,width=10cm}}
\caption{FIGURE 1. Plot of the (50--100 keV)/(20--50 keV) hardness ratio as a function of the 20--100 keV flux, 
determined 
from the BATSE data, using 3 day averages.  In red are
times of quiescent activity in the radio and in blue are times of flaring activity.}

\end{figure} 

\begin{figure}                                                                  
\centerline{\psfig{file=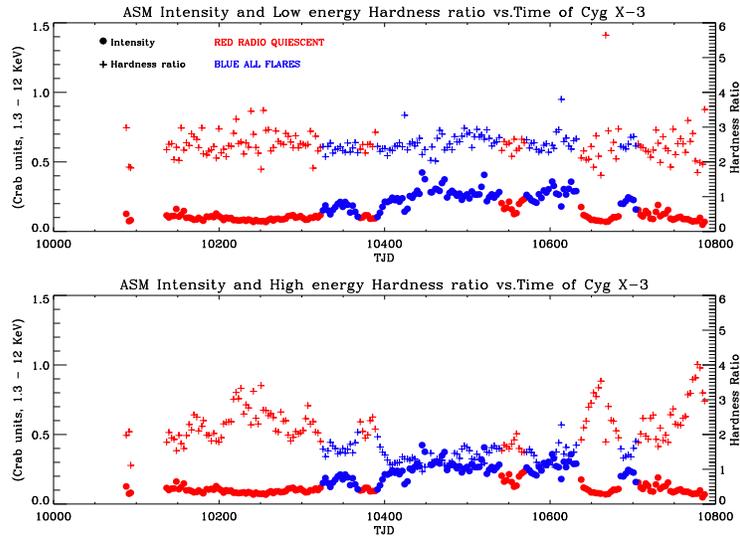,angle=90.0,height=7.25cm,width=10cm}}
\caption{FIGURE 2. {\it Top:} Plot of time histories of the ASM flux (1.3--12 keV) and the 
(3.0-4.8 keV)/(1.3--3.0 keV) hardness ratio, using 3 day averages.  The periods denoted in red
are periods of radio quiescence and periods denoted in blue are times of flaring activity. {\it Bottom:}
Plot of time histories of the ASM flux (1.3--12 keV) and the (4.8-12.0 keV)/(3.0--4.8 keV)
hardness ratio.} 

\end{figure} 



\begin{figure}                                                                  
\centerline{\psfig{file=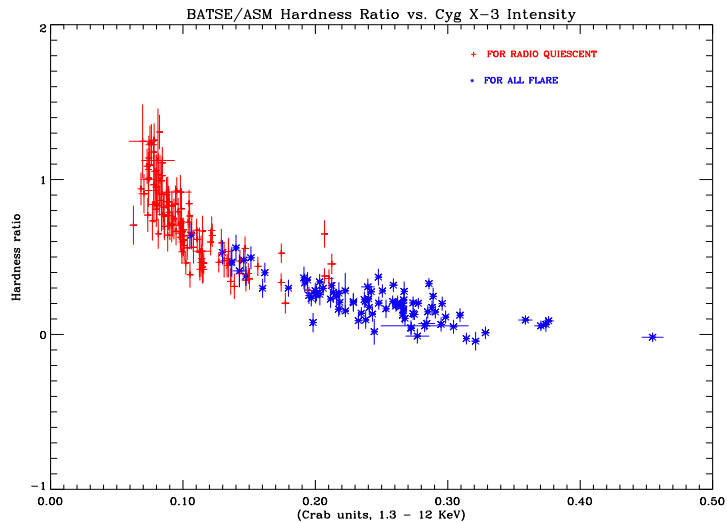,angle=90.0,height=7cm,width=10cm}}
\caption{FIGURE 3. Plot of the (20--50 keV)/(4.8--12 keV) hardness ratio determined from the BATSE and ASM 
data. Three day averages of the data were converted to Crab units and a hardness ratio was calculated.  The 
red are times of quiescent activity in the radio and the blue are times of flaring activity.}

\end{figure} 

\begin{figure}                                                                  
\centerline{\psfig{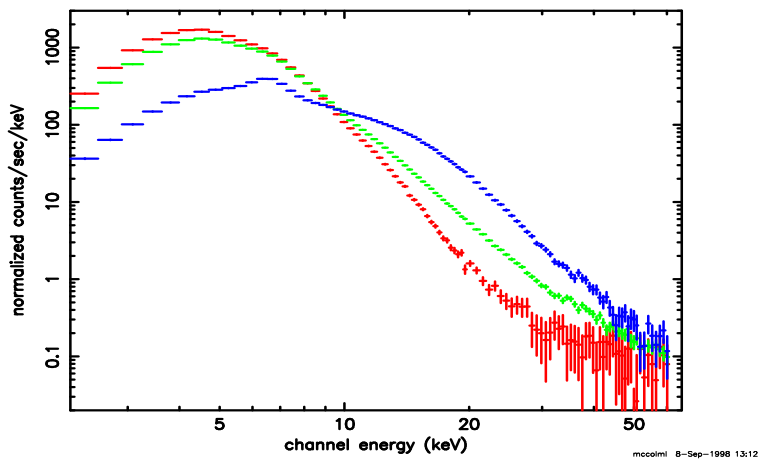}}
\caption{FIGURE 4. RXTE/PCA Count spectra are shown for three observations of Cyg X-3.  The spectrum
in red is the PCA observations taken just after the radio had come out of a state of quenched
emission.  The spectrum in green is the PCA observation taken just after a major radio flare.  The
spectrum in blue is the PCA observation taken during a radio quiescent state.  Each spectrum is a 500
second integration performed near the intensity peak of each observation.}

\end{figure}

\end{document}